\documentclass[11pt]{article}
\usepackage{moriond}

\usepackage{amsmath}
\usepackage{bm}
\usepackage{textcomp}

 \makeatletter
\g@addto@macro\bfseries{\boldmath}
\makeatother

\def\Journal#1#2#3#4{{#1} {\bf #2}, #3 (#4)}

\def\PLB{{\em Phys. Lett.}  B}
\def\PRL{\em Phys. Rev. Lett.}
\def\PRD{{\em Phys. Rev.} D}

\def\be{\begin{equation}}
\def\ee{\end{equation}}
\def\bea{\begin{eqnarray}}
\def\eea{\end{eqnarray}}

\newcommand{\Photo}{\includegraphics[height=35mm]{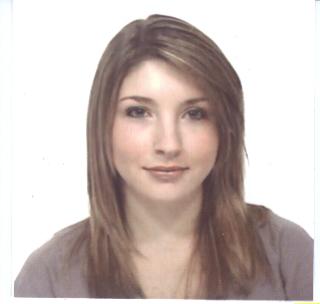}}

\begin{document}
\vspace*{4cm}
\title{$\mathbf{\textit{CP}}$ VIOLATION IN THE $\boldsymbol{B^0_{(s)}}$ SYSTEM AT LHCb}

\author{ F. DORDEI \\ on behalf of the LHCb Collaboration}

\address{Physikalisches Institut, Im Neuenheimer Feld 226, \\
69120 Heidelberg, Germany}

\maketitle\abstracts{
The study of $C\!P$ violation in decays of neutral \textit{B} mesons provides an important test of the Standard Model (SM) predictions and it is a sensitive probe to search for non-SM physics. In these proceedings I present measurements of $C\!P$ violation in the neutral \textit{B} meson system. The discussed analyses are based on $pp$-collision data corresponding to an integrated luminosity of 1 $\mathrm{fb}^{-1}$ or 3 $\mathrm{fb}^{-1}$ recorded by the LHCb experiment.
}

\section{Introduction}

The Standard Model description of $C\!P$ violation has been very successful in describing existing data.~\cite{PDG} However, the source of $C\!P$ violation in the SM is not sufficient~\cite{baryon} to explain the matter-antimatter asymmetry that results in the matter dominated universe we observe today. Therefore, precision measurements of $C\!P$ violation are mandatory in order to confirm SM predictions and to check whether non-SM contributions are also present.\\
\indent LHCb~\cite{LHCb} is one of the four large particle experiments located at the LHC. It is specialised for precision measurements of $b$- and $c$- hadron decays, searching for indirect effects of new virtual particles in quantum loops, predicted in New Physics models.
In particular decays of neutral $B$-mesons provide an ideal laboratory to study $C\!P$ violation originating from a non-trivial complex phase in the CKM quark mixing matrix~\cite{CKM}. Since the weak interaction eigenstates are not the same as the mass eigenstates, the flavour content of neutral mesons changes as a function of time, giving rise to the phenomenon known as neutral meson mixing. Thanks to this effect, neutral \textit{B} mesons allow to test all the three different kinds of $C\!P$ violation: $C\!P$ violation in the decay, $C\!P$ violation in the mixing and $C\!P$ violation in the interference between mixing and decay. The former is possible in both neutral and charged meson decays and it consists in a difference between absolute value of the amplitude of the decay of a meson to a certain final state and the amplitude of the $C\!P$ conjugated decay. The latter two are instead only possible for neutral mesons.\\
\indent The phenomenological aspects linked to neutral \textit{B} meson mixing are described in many articles~\cite{Nierste}; only the main parameters involved are briefly introduced here. $B^0_{(s)}$ and $\overline{B}^0_{(s)}$ are flavour eigenstates with the quark content $\overline{b}d(s)$ and $b\overline{d}(\overline{s})$, respectively. Any arbitrary combination of flavour eigenstates, $B^0_{(s)}$ and $\overline{B}^0_{(s)}$, has a time evolution described by an effective Schr{\"o}dinger equation (we adopt units such that $\hbar=\textit{c}=1$)
\begin{equation}\label{eq:Scheq}
\textit{i}\,\,\frac{d}{dt}\begin{pmatrix} B^0_{(s)} \\ \overline{B}^0_{(s)} \\ \end{pmatrix} = 
 \begin{pmatrix}
  M_{11}-i\,\Gamma_{11}/2 & M_{12}-i\,\Gamma_{12}/2  \\
  M_{21}-i\,\Gamma_{21}/2 & M_{22}-i\,\Gamma_{22}/2  \\
 \end{pmatrix} \begin{pmatrix} B^0_{(s)} \\ \overline{B}^0_{(s)} \\ \end{pmatrix}\,,
\end{equation}
where \textbf{M} and $\bm{\Gamma}$ are $2\times2$ hermitian matrices. The heavy (H) and light (L) mass eigenstates of the Schr{\"o}dinger equation are obtained diagonalizing the matrix in Eq.\ref{eq:Scheq}
\begin{equation}
 |B_{H/L}\rangle = p |{B}^0_{(s)}\rangle \pm q |\overline{B}^0_{(s)}\rangle \,,
\end{equation}
where the complex coefficients p and q obey the normalization condition $|p|^2+|q|^2=1$. The system can be defined in terms of five different experimental observables. They are the mass difference $\Delta m_{(s)}$ and the decay width difference $\Delta \Gamma_{(s)}$ between the mass eigenstates, defined by
\begin{equation}
\Delta m_{(s)} \equiv m_H - m_L \simeq 2|M_{12}|, \,\,\,\,\,\,\,\,\,\,\,\,\,\,\,\,\,\,\,\,\, \Delta \Gamma_{(s)} \equiv \Gamma_L - \Gamma_H \simeq 2 |\Gamma_{12}| \cos(\phi_{12}),
\end{equation}
the average mass and width, defined by
\begin{equation}
m_{(s)} = \frac{m_H+m_L}{2}, \,\,\,\,\,\,\,\,\,\,\,\,\,\,\,\,\,\,\,\,\, \Gamma_{(s)} = \frac{\Gamma_L+\Gamma_H}{2},
\end{equation}
and the so-called flavour-specific or semileptonic asymmetry, which is a dimensionless quantity that parametrizes $C\!P$ violation in the mixing and is approximately given by
\begin{equation}
  a_{sl} \simeq \frac{\Delta \Gamma}{\Delta M} \tan \phi_{12}\,.
\end{equation}
In the SM model the phase $\phi_{12}$ is very small, and in particular for $B^0_s$ mixing, $a^s_{sl}$ is predicted~\cite{LenzNierste} to be $(+1.9 \pm 0.3) \cdot 10^{-5}$.\\ 
\indent The interference between $B^0_s$-mesons decaying to a $J/\psi hh$ ($h$= $K$ or $\pi$) final state either directly or via $B^0_s-\overline{B}^0_s$ mixing gives rise to a $C\!P$-violating phase called $\phi_s$. In the SM, neglecting sub-leading penguin contributions, this phase is predicted~\cite{phis} to be $-2\,\, \mathrm{arg}(-V_{ts}V_{tb}^*/V_{cs}V_{cb}^*) = (3.40^{+1.32}_{-0.77})\times10^{-3}$, where $V_{ij}$ are elements of the CKM matrix. \\
\indent In the following a selection of recent LHCb measurements of $C\!P$ violation in the neutral \textit{B} meson system are presented. They are based on 2011 only or 2011+2012 data sets, corresponding to 1 $\mathrm{fb}^{-1}$ or 3 $\mathrm{fb}^{-1}$ of integrated luminosity of $pp$ collisions, respectively.
\section{Measurement of $\mathbf{\textit{CP}}$ asymmetries, and direct $\mathbf{\textit{CP}}$ violation in $B^0 \to \phi K^{*}(892)^{0}$}\label{sec:an1}
An angular analysis of the decay $B^0 \to \phi K^{*}(892)^{0}$ has been performed~\cite{phikstar} using a data set which corresponds to $1.0\,\,\mathrm{fb}^{-1}$ of integrated luminosity.  It is a pseudoscalar to vector-vector transition. For a P-wave state angular momentum conservation allows three possible helicity configurations of the vector meson pair, parametrised by a longitudinal polarisation $A_0$, and two tansverse polarisations, $A_\perp$ and $A_\parallel$. Moreover for the first time also the contributions where either the $KK$ or $K\pi$ are produced in an S-wave state are taken into account, with the amplitudes $A_s^{KK}$ and $A_s^{K\pi}$, respectively.\\
\indent The polarisation amplitudes are extracted by means of an untagged time-independent angular analysis. An almost equal mixture of longitudinal and transverse polarisations is found with greater precision than by previous measurements. The fraction of the longitudinal polarisation is measured to be $f_L = 0.497 \pm 0.019\,\,\mathrm{(stat.)} \pm 0.015\,\,\mathrm{(syst.)}$. Moreover, a significant $S$-wave contributions is found in both the $K\pi$ and the $KK$ system. Using the information that the \textit{B} meson flavour at decay can be identified by the charge of the kaon from the $K^*$, the sample is splitted in two sub-samples and the difference in polarisation amplitudes and phases between the two samples is calculated. All differences are found to be consistent with zero within the uncertainties.\\ \indent The polarisation amplitudes and phases are also used to calculate triple product asymmetries. Non-zero triple product asymmetries arise either due to a $T$-violating phase or a $C\!P$-violating 
phase and final-state interactions (so-called fake asymmetry). The former case (so-called true asymmetry) implies that $C\!P$ is violated, assuming $C\!PT$ is conserved. The true asymmetries are found consistent with zero, showing no evidence for $C\!P$ violation. In contrast, all but one of the fake asymmetries are significantly different from zero, indicating the presence of final-state interactions introducing strong phases.\\
\indent After splitting the sample according to the flavour of the \textit{B} meson at decay, the following raw asymmetry is measured
\begin{equation}
  A = \frac{N(\overline{B}^{0} \to \phi \overline{K}^*(892)^0)-N(B^0 \to \phi K^*(892)^0)}{N(\overline{B}^{0} \to \phi \overline{K}^*(892)^0)+N(B^0 \to \phi K^*(892)^0)}.
\end{equation}
Correcting for particle-antiparticle production and detection asymmetries, calculated using the control channel $B^0 \to J/\psi K^*(892)^0$, the direct $C\!P$ asymmetry is determined to be
\begin{equation}
  A_{C\!P}(\phi K^*(892)^0) = (+1.5 \pm 3.2 \, \mathrm{(stat.)} \pm 0.5 \, \mathrm{(syst.)}) \%.
\end{equation}
\indent This is consistent with zero and a factor of two more precise than the values reported by Babar and Belle~\cite{BabarBelle}.
\section{Measurement of time-dependent $\mathbf{\textit{CP}}$ violation in $B^0_s \to K^+K^-$ and $B^0 \to \pi^+\pi^-$}\label{subsec:an2}
The study of $C\!P$ violation in charmless charged two-body decays of neutral \textit{B} mesons is a good probe of the SM predictions. Moreover, precise measurements in this sector are important to constrain hadronic factors that cannot be accurately calculated from quantum chromodynamics at present. In these proceedings the first measurement of time-dependent $C\!P$-violating asymmetries in $B^0_s \to K^+K^-$ decays~\cite{BKK} is discussed. Furthermore, a measurement of the corresponding quantities for $B^0 \to \pi^+\pi^-$ has been performed. The analysis is based on a data sample corresponding to $1\,\,\mathrm{fb}^{-1}$ of integrated luminosity.  \\
\indent Assuming $C\!PT$ invariance, the $C\!P$ asymmetry as a function of time for neutral \textit{B} mesons decaying to a $C\!P$ eigenstate $f$ is given by
\begin{equation}
A^{\mathrm{C\!P}}(t) =\frac{\Gamma_{\overline{B}^0_{(s)}\to f}(t) - \Gamma_{B^0_{(s)} \to f}(t)}{\Gamma_{\overline{B}^0_{(s)}\to f}(t) +  \Gamma_{B^0_{(s)} \to f}(t)} = \frac{-C_{f} \cos(\Delta m_{(s)}t) + S_{f} \sin(\Delta m_{(s)}t)}{\cosh \left( \frac{\Delta \Gamma_{(s)}}{2}t \right) - \mathcal{A}^{\Delta \Gamma_{(s)}}_{f} \sinh \left( \frac{\Delta \Gamma_{(s)}}{2}t \right)}\,.
\end{equation}
The quantities $C_f$, $S_f$, and $\mathcal{A}^{\Delta \Gamma_{(s)}}_f$ are
\begin{equation}\label{adeltagamma}
  C_f = \frac{1-|\lambda_f|^2}{1+|\lambda_f|^2}, \,\,\,\,\,\,\,\,\,\,\,\, S_f = \frac{2 \mathrm{Im} \lambda_f}{1+|\lambda_f|^2}\,,  \,\,\,\,\,\,\,\,\,\,\,\, \mathrm{and}\,\,\,\,\,\,\,\,\,\,\,\, \mathcal{A}^{\Delta \Gamma_{(s)}}_f=-\frac{2\mathrm{Re}\lambda_f}{1+|\lambda_f|^2}\,,
\end{equation}
where the parameter $\lambda_f=(q\,\overline{A}_f)/(p\,A_f)$ is related to the $B^0_{(s)}$ meson mixing (via $q/p$) and to the decay  amplitudes of the $B^0_{(s)} \to f$ decay ($A_f$) and of the $\overline{B}^0_{(s)} \to f$ decay ($\overline{A}_f$).\\ 
The parameter $\mathcal{A}^{\Delta \Gamma_{(s)}}_f$ can be expressed as $\mathcal{A}^{\Delta \Gamma}_f = \pm \sqrt{1-C_f^2-S_f^2}$.
For $B^0_s$ decays the positive solution is taken, which is consistent with measurement of the $B^0_s \to K^+ K^-$ effective lifetime, while for $B^0$ decays, due to the fact that the width difference of the $B^0$ meson is negligible, the ambiguity is not relevant. By means of a time-dependent analysis, where the initial $B^0_{(s)}$ flavour is identified through a flavour-tagging algorithm calibrated using flavour-specific $B^0 \to K^+ \pi^-$ events, the parameters for $B^0_s \to K^+ K^-$ decay are found to be
\begin{eqnarray}
  C_{KK} = 0.14 \pm 0.11\,\mathrm{(stat.)} \pm 0.03 \,\mathrm{(syst.)},\\ \nonumber
  S_{KK} = 0.30 \pm 0.12\,\mathrm{(stat.)} \pm 0.04 \,\mathrm{(syst.)},
\end{eqnarray}
with a statistical correlation coefficient of 0.02. The results for the $B^0 \to \pi^+\pi^-$ decay are
\begin{eqnarray}
  C_{\pi\pi} = -0.38 \pm 0.15\,\mathrm{(stat.)} \pm 0.02 \,\mathrm{(syst.)},\\ \nonumber
  S_{\pi\pi} = -0.71 \pm 0.13\,\mathrm{(stat.)} \pm 0.02 \,\mathrm{(syst.)},
\end{eqnarray}
with a statistical correlation coefficient of 0.38. The significances for $(C_{KK},\,S_{KK})$ to differ from $(0,0)$ are determined to be 2.7$\,\sigma$ and 5.6$\,\sigma$, respectively. These results are compatible with SM predictions and are important inputs in the determination of the unitarity triangle angle $\gamma$ using decays affected by penguin processes.
\section{Measurement of the semileptonic asymmetry $a^{s}_{sl}$}\label{subsec:an3}
The LHCb collaboration has performed a measurement~\cite{asl_meas} of the semileptonic asymmetry $a^{s}_{sl}$ in decays of $B^0_s\to D^+_s \mu^-X$, using a data sample corresponding to an integrated luminosity of $1 \,\, \mathrm{fb}^{-1}$. The time integrated asymmetry in the untagged yields of $D_s^- \mu^+$ and $D^+_s \mu^-$ events is related to $a_{sl}^s$ by
\begin{equation} \label{eq:als}
A^{\mathrm{C\!P}}_{\mathrm{measured}} = \frac{\Gamma [D^{-}_{s} \mu^{+}]-\Gamma [D^{+}_{s} \mu^{-}]}{\Gamma [D^{-}_{s} \mu^{+}]+ \Gamma [D^{+}_{s} \mu^{-}]} = \frac{a^{s}_{sl}}{2}+ a_D +\left[ a_p - \frac{a^{s}_{sl}}{2}\right] \cdot \frac{\int e^{-\Gamma_s t}\cos(\Delta m_s t) \varepsilon(t) dt}{\int e^{-\Gamma_s t}\cosh(\Delta \Gamma_s/ 2t) \varepsilon(t) dt},
\end{equation}
where $\varepsilon(t)$ is the decay time acceptance function and $a_{p}$ is the particle-antiparticle production asymmetry. The latter is expected to be at most a few percent at the LHC. However, thanks to the large value of the oscillation frequency~\cite{deltam}, $\Delta m_s = 17.768 \pm 0.024\,\mathrm{ps}^{-1}$, the integral on the right-hand side is small and the third term on the right-hand side of Eq.\ref{eq:als} can be neglected. \\
\indent Using data-driven methods to measure efficiency ratios, the measured $C\!P$ asymmetry is corrected for tracking efficiency asymmetries and background asymmetries ($a_D$) and the semileptonic asymmetry $a^s_{sl} = (-0.06 \pm 0.50 \pm 0.36)\%$ is derived. This result is the most precise measurement to date and is consistent with the SM.
\section{Measurement of the $\mathbf{\textit{CP}}$-violating phase $\phi_s$}\label{subsec:an4}
The decays $B^0_s \rightarrow J/\psi \phi$ and $B^0_s \rightarrow J/\psi \pi^+\pi^-$ are used to measure the $C\!P$ violating phase, $\phi_s$. The decay $B^0_s \rightarrow J/\psi \phi$ is a pseudo-scalar to vector-vector decay, so angular momentum conservation implies that the final state is an admixture of $C\!P$-even and $C\!P$-odd components. By performing a tagged time-dependent angular analysis, it is possible to statistically disentangle the different $C\!P$ eigenstates by the differential decay rate for $B^0_s$ and $\overline{B}^0_s$ mesons, produced as flavour eigenstates at $t=0$. Using $1$ $\mathrm{fb}^{-1}$ of data collected in $pp$ collisions at $\sqrt{s}=7$ TeV with the LHCb detector, the $C\!P$ violating phase $\phi_s$ as well as $\Delta \Gamma_s$ and $\Gamma_s$ are extracted~\cite{phis_meas} by performing an unbinned maximum log-likelihood fit to the $B^0_s$ mass, decay time $t$, angular distributions and initial flavour tag of the selected $B^0_s \rightarrow J/\psi \phi$ events.
The $B^0_s \rightarrow J/\psi \pi^+\pi^-$ final state is predominantly $C\!P$-odd. Thus there is no need for an angular analysis. The results of a simultaneous fit to both $B^0_s \rightarrow J/\psi \phi$ and $B^0_s \rightarrow J/\psi \pi^+\pi^-$ are: $\phi_s = 0.01 \pm 0.07 \,\mathrm{(stat.)} \pm 0.01 \,\mathrm{(syst.)}$, $\Gamma_s = (0.661 \pm 0.004 \,\mathrm{(stat.)} \pm 0.006 \,\mathrm{(syst.)})\,\mathrm{ps}^{-1}$ and $\Delta \Gamma_s = (0.106 \pm  0.011 \, \mathrm{(stat.)} \pm 0.007 \,\mathrm{(syst.)}) \,\mathrm{ps}^{-1} \label{eq:deltagamma}$.\\
\indent These results are in good agreement with the SM prediction. The systematic uncertainty on $\phi_s$ is dominated by the size of the Monte Carlo sample used to determine the angular acceptance, the one on $\Delta \Gamma_s$ is dominated by the background subtraction method and the decay time acceptance, the one on $\Gamma_s$ by the decay time acceptance determination. All of them are expected to decrease in the near future.
\section{Measurement of resonant and $\mathbf{\textit{CP}}$ components in $\overline{B}^0_s \to J/\psi\,\,\pi^+\pi^-$ decays}\label{subsec:an5}
Using the dataset corresponding to $3$ $\mathrm{fb}^{-1}$ of integrated luminosity, the resonant structure of the decay $\overline{B}^0_s \to J/\psi\,\,\pi^+\pi^-$ has been studied~\cite{pipi}. The $\pi^+\pi^-$ invariant mass and all three decay angular distributions are used to determine the resonant and non-resonant components. The decay $\overline{B}^0_s \to J/\psi\,\,\pi^+\pi^-$ can be described by the interfering sum of five resonant components: $f_0(980)$, $f_0(1500)$, $f_0(1790)$, $f_2(1270)$ and $f^\prime_2(1525)$. An alternative model including these states and a non-resonant $J/\psi \pi^+ \pi^-$ component has been checked and it also provides a good description of data. In both models the largest component of the decay is $f_0(980)$, while the $f_0(500)$ and the $\rho(770)$ components are found to be not signficant. An upper limit for the fit fractions of these two component is derived, corresponding to $3.4\%$ and $1.7\%$ at 90$\%$ confidence level (from the solution where also the non-resonant 
component is included), respectively.\\
\indent The final state is found to be compatible with being entirely $C\!P$-odd; the $C\!P$-even part is measured to be less than 2.3$\%$ at 95$\%$ CL.  This upper limit is the same as of a previous measurement~\cite{pipi_old} performed with $1$ $\mathrm{fb}^{-1}$ of integrated luminosity, while the current measurement also adds a possible $f^\prime_2(1525)$ contribution.\\
\indent Also of importance is the limit on the absolute value of the mixing angle, $\phi_m$, between the $f_0(500)$ and the $f_0(980)$ resonances. The limit is found to be
\begin{equation}
  |\phi_m| < 7.7^\circ\,\,\,\,\, \mathrm{at}\,\, 90\%\,\, \mathrm{CL},
\end{equation}
and it is the most constraining ever placed. It is consistent with the tetraquark model, which predicts zero within a few degrees.
\section{Measurement of the $\overline{B}^0_s \to D^+_s D^-_s$ effective lifetime}\label{subsec:an6}
Measurements of the $B_s^0$ effective lifetime in decays to $C\!P$-odd and $C\!P$-even flavour specific final states, $f$, allow to probe the decay width difference, $\Delta \Gamma_s$, and the $C\!P$-violating phase, $\phi_s$, of $B^0_s-\overline{B}^0_s$ mixing box-diagram.
We define the effective lifetime of the decay $B^0_s \rightarrow f$ as the time expectation value of the untagged rate
\begin{equation}
  \tau_f \equiv \frac{\int_0^\infty t \,\,\langle \Gamma (B^0_s(\overline{B}^0_s)(t) \rightarrow f) \rangle dt}{\int_0^\infty \langle \Gamma (B^0_s(\overline{B}^0_s)(t) \rightarrow f) \rangle dt}\,,
\end{equation}
which is equivalent to the lifetime that results from fitting the untagged decay time distribution with a single exponential. 
By making use of the usual definition $y_s = \Delta \Gamma_s / 2\Gamma_s$ and using $\tau_{B^0_s}= \Gamma_s^{-1}$, one can express the effective lifetime as
\begin{equation}
\tau_f = \tau_{B^0_s} + \tau_{B^0_s}\, \mathcal{A}^{\Delta \Gamma_{s}}_f\, y_s + \tau_{B^0_s}\, [2-(\mathcal{A}^{\Delta \Gamma_{s}}_f)^2]\, y_s^2 + \mathcal{O}(y_s^3)\,.
\end{equation}
The parameter $\mathcal{A}^{\Delta \Gamma_{s}}_f$ is defined in Eq. \ref{adeltagamma}, and is a function of $\phi_s$~\cite{Fleischer}.
So the corresponding lifetime measurements can be used to constrain $y_s$ (or $\Delta \Gamma_s$) with respect to $\phi_s$, with the advantage that only an untagged analysis is needed.\\
\indent  Here the first measurement of the effective lifetime of the $\overline{B}^0_s$ meson in the decay $\overline{B}^0_s \to D^+_s D^-_s$ is reported~\cite{lifetime}, using a dataset corresponding to $3\,\,\mathrm{fb}^{-1}$ of integrated luminosity. The lifetime of $\overline{B}^0_s \to D^+_s D^-_s$ meson is measured relative to the well known $B^-$ lifetime, using the normalisation decay channel $B^- \to D^0 D^-_s$, which has a similar topology and kinematic properties. As a result, many of the systematic uncertainties cancel in the ratio.
The measured value of the $\overline{B}^0_s \to D^+_s D^-_s$ effective lifetime is $\tau = (1.379 \pm 0.026 \,\mathrm{(stat.)} \pm 0.017 \,\mathrm{(syst.))}$ ps. The $D^+_s D^-_s$ final state is $C\!P$-even hence the effective lifetime is approximately equal to $\Gamma^{-1}_L$. Using this fact a value of $\Gamma_L = (1.52 \pm 0.15 \,\mathrm{(stat.)} \pm 0.001 \,\mathrm{(syst.)})\,\,\mathrm{ps}^{-1}$ is derived.
\section{Conclusions}
Recent LHCb results of $C\!P$ violation in the neutral \textit{B} meson system have been presented. All measurements are in good agreement with SM predictions. The selection of $C\!P$ violation presented here is not a complete list of the analyses performed by the LHCb collaration. Updated results based on the full $3\,\,\mathrm{fb}^{-1}$ of data are expected soon and will allow to constrain even stronger contributions of beyond the SM physics.

\section*{Acknowledgments}
The author acknowledges the support received from the ERC under FP7 and the support by the International Max Planck Research School for Precision Tests of Fundamental Symmetries.

\end{document}